\documentclass[fleqn,usenatbib]{mnras}

\usepackage{newtxtext,newtxmath}

\usepackage[T1]{fontenc}
\DeclareRobustCommand{\VAN}[3]{#2}
\let\VANthebibliography\thebibliography
\def\thebibliography{\DeclareRobustCommand{\VAN}[3]{##3}\VANthebibliography}

\usepackage{graphicx}	
\usepackage{amsmath}	
\usepackage{subcaption}
\usepackage{float}
\usepackage[export]{adjustbox}

\title[]{Experimental phase function and degree of linear polarization curve of olivine and spinel and the origin of the Barbarian polarization behavior}

\author[E. Frattin et al.]{E. Frattin,$^{1,2}$
 J. Martikainen,$^{1,3}$
 O. Mu\~{n}oz,$^{1}$\thanks{E-mail: olga@iaa.es} 
  J. C. G\'{o}mez-Mart\'{i}n,$^{1}$
 T. Jardiel,$^{4}$\newauthor
 A. Cellino, $^{5}$ 
 G. Libourel,$^{6}$
 K. Muinonen,$^{3}$  
 M. Peiteado,$^{4}$ 
  P. Tanga,$^{6}$  \newauthor
  \\      
   $^{1}$  Instituto de Astrof\'{\i}sica de Andaluc\'{\i}a, CSIC, Glorieta de la Astronomia s/n, E-18008 Granada, Spain\\
    $^{2}$  Dipartimento di Fisica e Astronomia ‘G. Galilei’, University of Padova, Vicolo dell’Osservatorio
3, I-35122 Padova, Italy.\\
    $^{3}$ Department of Physics, University of Helsinki, Finland\\
$^{4}$  Instituto de Cer\'{a}mica y Vidrio, CSIC, C/ Kelsen 5, Campus Cantoblanco, 28049 Madrid, Spain.
\\
 $^{5}$  INAF – Astrophysical Observatory of Torino, Via Osservatorio 20, I-10025 Pino Torinese (TO),
Italy. \\
  $^{6}$  Universit\'{e} Cote d’Azur, Observatoire de la Cote d’Azur, CNRS, Laboratoire Lagrange UMR7293, Nice, France.\\
     }

\date{Accepted XXX. Received YYY; in original form ZZZ}

\pubyear{2022}

\begin{document} 
\label{firstpage}
\pagerange{\pageref{firstpage}--\pageref{lastpage}}
\maketitle
\begin{abstract}
We explore experimentally possible explanations of the polarization curves of the sunlight reflected by the Barbarian asteroids. Their peculiar polarization curves are characterized by a large inversion angle, around 30$^{\circ}$, which could be related to the presence of FeO-bearing spinel embedded in Calcium-Aluminum Inclusions. 
In order to test this hypothesis, we have measured the phase function and  degree of linear polarization of six samples of Mg-rich olivine and spinel. 
For each material, we have analyzed the light scattering properties of a millimeter-sized grain and of two powdered samples with 
 size distributions in the micrometer size range.  
The three spinel samples show a well-defined negative polarization branch with an inversion phase angle located around 24$^{\circ}$-30$^{\circ}$. In contrast,  in the case of the olivine samples, the inversion angle is highly dependent on particle size and tends to decrease for  larger sizes. 
We identify the macroscopic geometries as a possible explanation for the evident differences in the polarization curves between olivine and spinel millimeter  samples. 
Although the polarization behaviour in near backscattering of the Barbara asteroid is similar to that of our spinel mm-sized sample in random orientation,  this similarity could result in part from crystal retro-reflection rather than composition.
This is part of an ongoing experimental project devoted to test separately several components of CV3-like meteorites, representative of the Barbarians composition, to disentangle their contributions to the  polarization behavior of these objects.

\end{abstract}




\section{Introduction}
Polarimetric observations are a powerful tool to understand the nature of   asteroids. The relations among the polarization curve parameters and the spectral behavior of asteroids help to refine their taxonomic classification \citep{fornasier2006, belskaya2017, lopez-sisterna2019} and are useful to identify those having a possible cometary origin \citep{cellino2018}.
Some asteroids have been found to share a peculiar polarimetric behavior and are commonly named Barbarians, after the prototype of this class (234) Barbara \citep{cellino2006,cellino2014,cellino2019,masiero2009,devogele2018}.
They are thought to be the remnants of a  generation of planetesimals accreted in the first epoch of Solar System formation \citep[and reference therein]{cellino2014}.
These asteroids are characterized by a polarization curve with an extended negative branch and large inversion angle, located at a phase angle of around 30$^{\circ}$.\\
Several assumptions have been proposed  to explain the large inversion angles of the Barbarian asteroids. 

Spectroscopical studies of the Barbarians have shown a characteristic absorption feature around 1 and 2 microns  related to the presence of spinel \citep{sunshine2008,devogele2018}.
Spinel is a  mineral component of   CAIs (Calcium Aluminum Inclusions), likely the oldest solid matter   in the Solar System and commonly found in all types of chondritic meteorites \citep{amelin2002}.
The 2-micron absorption band in the CAIs of the 'fluffy' type   is associated with the presence of spinel \citep{connolly2006}. The strength of the absorption band is determined primarily by the FeO contained in spinel \citep{sunshine2008}. 
Further, \cite{devogele2018} found that all Barbarians they analyzed belong to the  L-type taxonomic class as defined by \cite{demeo2009} and that their spectra can be modeled considering CAIs, olivine, and the typical  mineral compounds found in  CV3 meteorites. \\

In this paper, we present the experimental phase function and the degree of linear polarization (DLP) of  micron-sized and millimeter-sized samples of olivine and spinel.
This is the first part of an ongoing experimental project that aims to  disentangle the contributions of several mineral components of the Barbarians to their polarization behavior.

The paper is organized as follows: Section \ref{experimental_apparatus} is devoted to the description of the  experimental apparatus.
In Section~\ref{samples}, we describe the samples and in Section~\ref{measurements} we present the experimental results. 
Section~\ref{observations} discusses the relevance of our laboratory measurements for the interpretation of photo-polarimetric observations of the Barbarian asteroids. 
Finally, Section~\ref{conclusions} summarizes our conclusions.

\begin{figure*}
\begin{subfigure}{.45\textwidth}
    \centering
    \includegraphics[width = 0.8\linewidth]{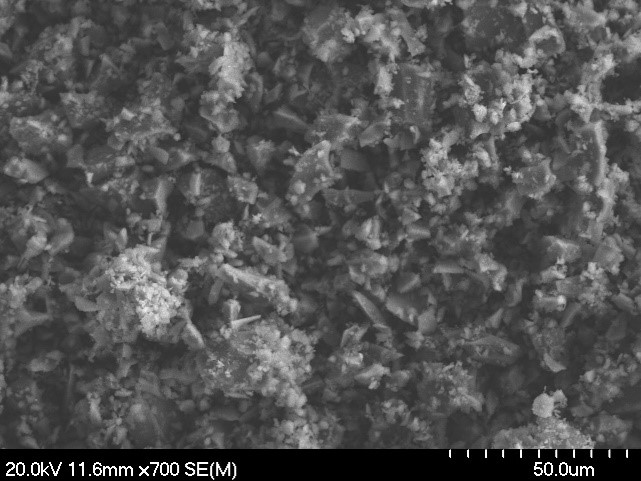}
    \caption{}
     \label{a}
\end{subfigure}
\begin{subfigure}{.45\textwidth}
\centering
     \includegraphics[width =  0.8\linewidth]{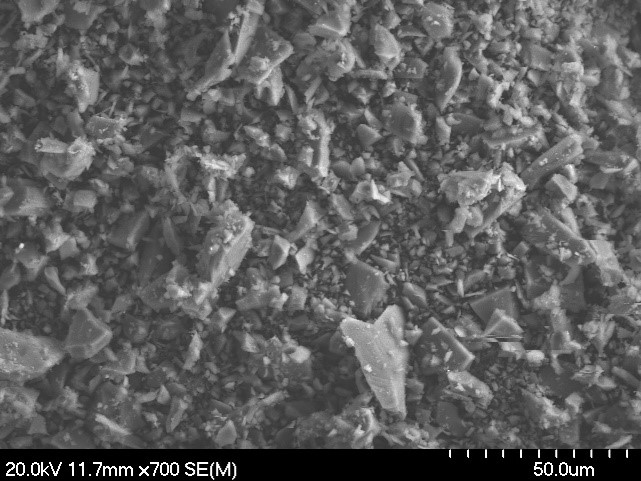}
 \caption{ }
     \label{b}
     \end{subfigure}
\begin{subfigure}{.45\textwidth}
    \centering
    \includegraphics[width = 0.8\linewidth]{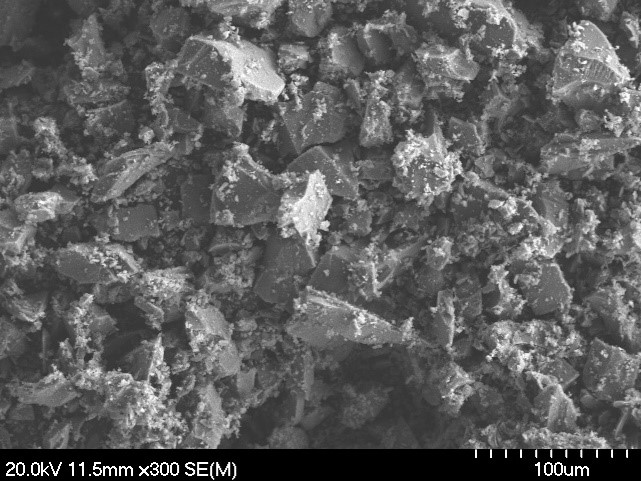}
    \caption{}
     \label{c}
\end{subfigure}
\begin{subfigure}{.45\textwidth}
    \centering
    \includegraphics[width = 0.8\linewidth]{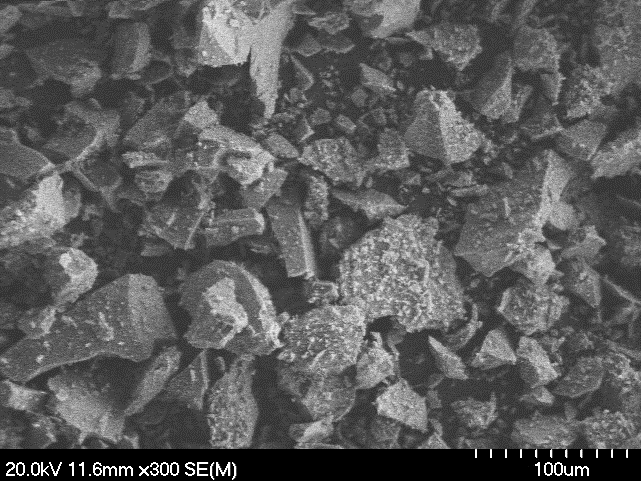}
    \caption{}
    \label{d}
\end{subfigure}
\begin{subfigure}{.45\textwidth}
    \centering
    \includegraphics[width = 0.8\linewidth]{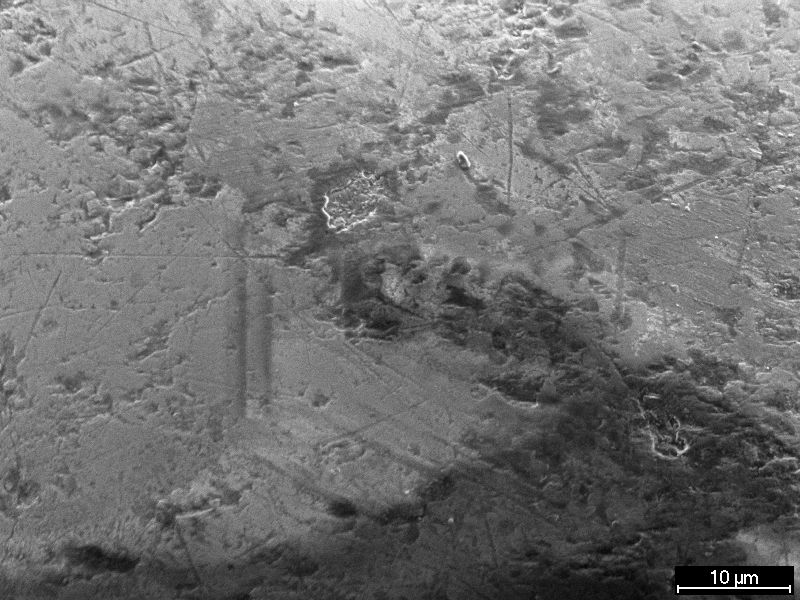}
    \caption{}
     \label{e}
\end{subfigure}
\begin{subfigure}{.45\textwidth}
    \centering
    \includegraphics[width = 0.8\linewidth]{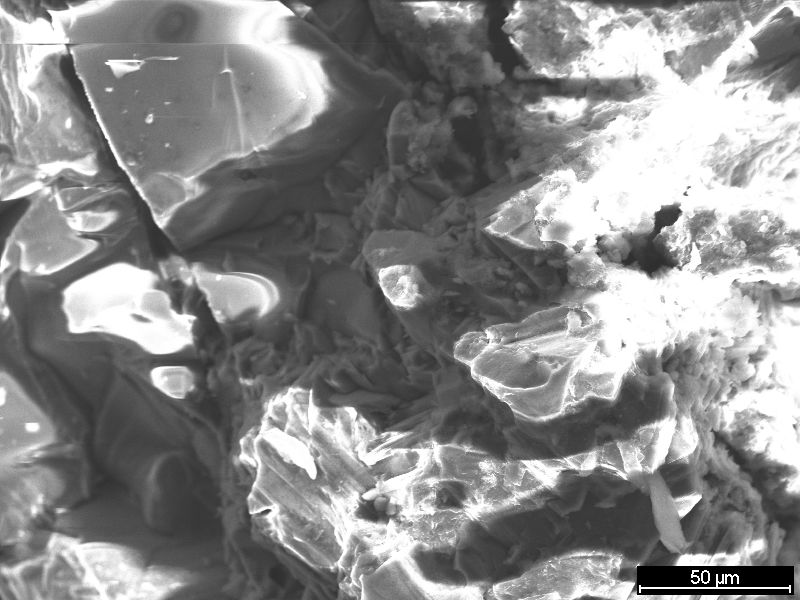}
    \caption{}
     \label{f}
\end{subfigure}
\begin{subfigure}{.45\textwidth}
    \centering
    \includegraphics[width = 0.8\linewidth]{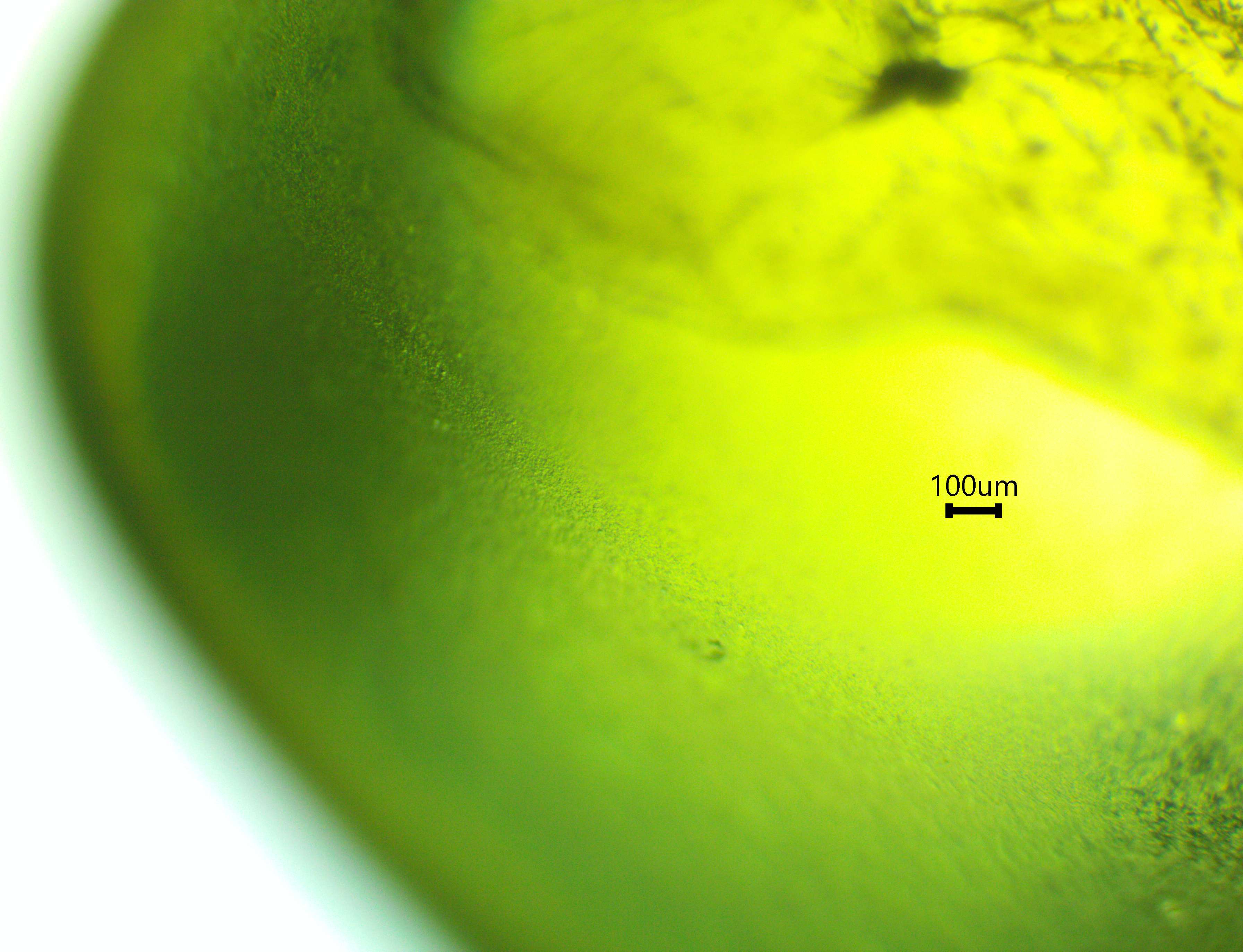}
    \caption{}
     \label{g}
\end{subfigure}
\begin{subfigure}{.45\textwidth}
    \centering
    \includegraphics[width = 0.8\linewidth]{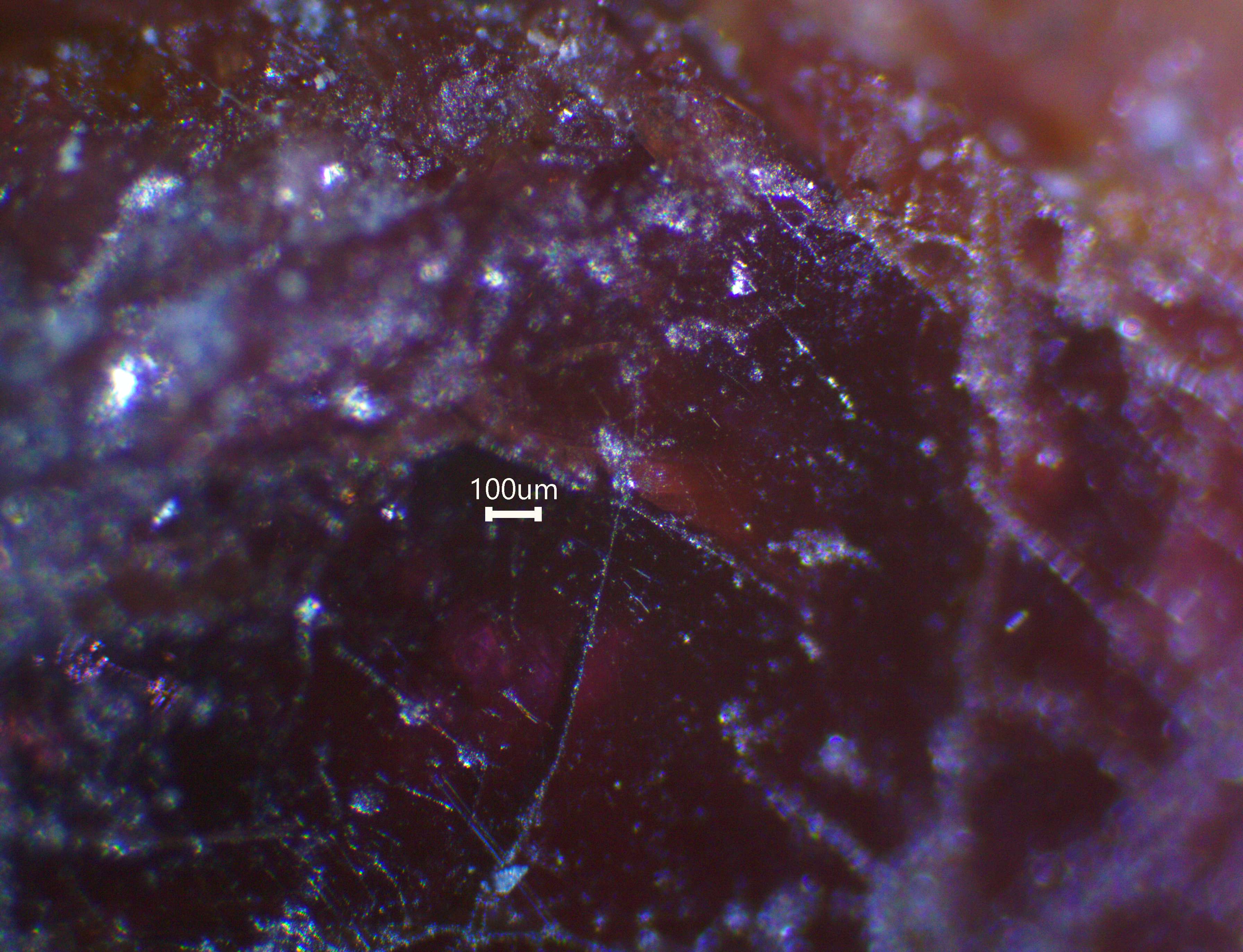}
    \caption{}
     \label{h}
\end{subfigure}
     \caption{FSEM micro-photographs of the \textit{Olivine Small} (a), \textit{Spinel Small} (b),  \textit{Olivine Medium} (c), \textit{Spinel Medium} (d), \textit{Olivine Pebble} (e), \textit{Spinel Pebble} (f). Optical microscopy images of the \textit{Olivine Pebble} and the \textit{Spinel Pebble} are also shown in panels g and h, respectively.     
    }
     \label{fig:images}
\end{figure*}

\section{Experimental apparatus}
\label{experimental_apparatus}
The measurements reported in this paper have been carried out at the IAA Cosmic Dust Laboratory (CODULAB) described in \cite{olga2011}. Briefly, a light beam generated by a diode laser that emits at 514 nm
passes through a polarizer and an electro-optic modulator and is subsequently scattered by a cloud of randomly oriented dust particles produced by  an aerosol generator. 
The scattered light is detected by a photomultiplier tube that moves around a 1-m diameter ring. Another photomultiplier tube, the Monitor, is located in a fixed position. The signal of the Monitor is used to correct from fluctuations in the aerosol beam and/or the incident laser beam. We combine polarization modulation of the incident beam and lock-in detection to determine per measurement run three elements (or combination of elements), $\rm F_{\rm i,j}$, of the 4$\times$4 scattering matrix, {\bf F}, of the dust sample.  The scattering matrix elements are functions of the number and physical properties of the scattering particles (size, morphology and refractive index), wavelength of the incident radiation, and scattering direction.  We refer to \cite{hulst} and \cite{hovenier}, for detailed description of the scattering matrix formalism. When the cloud consists of randomly oriented particles and time reciprocity applies, as is the case in our experiment, all scattering planes are equivalent and the scattering direction is fully described by means of the scattering angle, $\rm \theta$. Further, when the incident light is unpolarized the -$\rm F_{\rm 12}(\theta)$/$\rm F_{\rm 11}(\theta)$ ratio is equal to the degree of linear polarization, hereafter DLP. In our experiment the values of the $\rm F_{\rm 11}(\theta)$ element are normalized to 1 at $\theta$=30$^{\circ}$. The $\rm F_{\rm 11}(\rm \theta)$, normalized in this way,  is proportional to the flux of the scattered light when the incident light is unpolarized and is called phase function in this paper. To facilitate a direct comparison with the astronomical observations we use the phase angle $\alpha = 180^{\circ} - \theta$  throughout the text.

The apparatus performance has been tested  by comparing the measured scattering matrix of a  cloud of water droplets generated in situ by a nebulizer with  Mie computations for a distribution of homogeneous spherical  particles \citep{olga2010}.  \\
To measure the  scattering matrix of mm-sized single particles, the experimental apparatus has been modified as described in \cite{olga2020}. In these experiments the light source is an Argon-Kripton laser tuned at 520 nm.  A single particle is positioned on a rotary conical holder able to rotate \citep{olga2017, olga2020}. To simulate  random orientation, the $F_{xy}(\theta)$ is the result of averaging over 54 $F_{xy}^{\varphi}$ corresponding to 54 different orientations ($\varphi$) of the particle. The measurements are taken by rotating the holder  360$^{\circ}$ in steps of 10$^{\circ}$ around the vertical axis. Then, the particle is rotated  90$^{\circ}$ on the holder toward the direction of the laser beam and additional measurements are taken by rotating the holder  360$^{\circ}$ in steps of 20$^{\circ}$ around its vertical axis. The final value of the phase function and DLP at each phase angle is the average of the measurements at the 54 different positions of the sample. 

\section{Samples Description}
\label{samples}
    
\subsection{Chemical composition}
\label{composition}
In this work, we study samples of micron-scale particles and mm-sized pebbles of two types of material: olivine, a magnesium-iron silicate, and spinel, a magnesium-aluminium oxide.
We consider three sample sizes for each material: 
a millimeter-sized particle, hereafter labelled as pebble following the nomenclature for cometary dust as  described in \cite{guttler2019} and two  powdered samples of micron-scale particles.
The \textit{Olivine Pebble} and \textit{Spinel Pebble} were directly selected from the original coarse-grained material available.
The powdered samples were obtained by mechanical milling of some of the original coarse-grained material for five minutes. 
The resulting powder was sifted through 63 and 20 $\mu$m sieves in order to generate four samples named \textit{Olivine  Medium}, \textit{Olivine Small}, \textit{Spinel Medium}, and \textit{Spinel Small}.\\
HRTEM-EDX (High-Resolution Transmission Electron Microscope - Energy Dispersive X-Ray) micro analysis of the powdered samples revealed that the olivine   is magnesium rich, close to the endmember forsterite, with chemical composition Fe$_{0.16}$Mg$_{1.84}$SiO$_4$ (n=11), i.e. close to a classical San Carlos Fo90 olivine \citep{jarosewich1980}.
The spinel sample has a chemical composition (Mg$_{0.96}$Fe$_{0.04}$)$_{0.01}$(Al$_{1.97}$Cr$_0.03$)$_2$O$_4$ (n=6), close to the Vietnam pink and red spinels surveyed by \cite{giuliani2017} that we take as reference. 
The chemical  compositions of the two minerals are summarized in Tables \ref{tab:elemental_composition_olivine} and \ref{tab:elemental_composition_spinel}.\\

\begin{table}
\centering
\begin{tabular}{ccc}
\hline
&Olivine& \\
 \hline
FeO & 9.55 & 8  \\
MgO & 49.42 & 51  \\
SiO$_2$ & 40.81 & 41\\
CaO & < 0.05 & -  \\
MnO & 0.14 & -  \\
NiO & - & <0.1 \\
& \citet{jarosewich1980} & this work \\
\hline
\end{tabular}
\caption{Chemical analysis of a reference San Carlos olivine in wt\%  by \citet{jarosewich1980} (second column) and of the sample analyzed  in this work (third column).}
\label{tab:elemental_composition_olivine}
\end{table}
\begin{table}
     \centering
     \begin{tabular}{ccc}
\hline
&Spinel&\\
 \hline
FeO & <1 & 2 \\
MgO & 28 & 30 \\
Al$_2$O$_3$ & 70 & 66\\
TiO$_2$ & <0.02 & - \\
 V$_2$O$_3$ & <0.5 & - \\
 Cr$_2$O$_3$ & <1 & 1 \\
& \citet{giuliani2017} & this work \\
\hline
     \end{tabular}
     \caption{Chemical analysis of a pink and red Vietnam spinels  in wt\%   by \citet{giuliani2017} (second column) and of the red spinel sample studied in this work (third column). }
     \label{tab:elemental_composition_spinel}
 \end{table}

\subsection{Morphology}
\label{morphology}
Figure \ref{fig:images} shows the Field Emission Scanning Electron Microscope (FESEM) of the olivine and spinel powdered samples, which consist of irregular dust particles with sharp edges.
Both \textit{Medium} samples show high surface roughness as a result of the presence of micron-sized particles adhered to the surface of the larger particles.
The \textit{Small} samples also have a similar surface structure, although, olivine has a slightly more agglomerated structure than spinel.\\
The \textit{Pebble} samples are semi-transparent compact natural mm-sized particles.
Optical microscopy images are shown in Figures \ref{g} and \ref{h}.
Olivine crystallizes in the orthorhombic system, but our  sample is rounded, possibly due to alluvial wear. Spinel belongs to the cubic crystal system, and our sample shows a typical crystal habit with sharp angles and flat faces.
The apparently high level of roughness of \textit{Spinel Pebble} in Figure \ref{f} is due to the transparency of the material that allows the electron beam to reach different depth levels.

\subsection{Refractive index}
\label{refractive_index}
The refractive index of a medium is defined as:
\begin{equation}
m = c\sqrt{\epsilon \mu} = n + ik
\label{eq:index}
\end{equation}
where $\epsilon$ is the electric permittivity, $\mu$ is the magnetic permeability and $c$ the speed of light in vacuum. 
The optical constants $n$ represents the phase velocity of the wave in the medium, while $k$ is the absorption coefficient of the material. \\
An estimate of the refractive index at 514 nm  of the Mg-rich olivine samples used in this work is $m = 1.62 + i10^{-5}$. This number has been obtained from the Jena-St. Petersburg Database of Optical Constants (http://www.astro.uni-jena.de/Laboratory/Database/jpdoc/index.html).
An estimate of the refractive index of the spinel samples at 500 nm  is $m = 1.72 + i3\times10^{-4}$  \citep{spinel_refractive_index, zeidler2011}.
The imaginary part of the refractive indexes is quite small, indicating a low absorbance of these materials. The transmission through the particle is strongly damped for the pebbles as k*2x (where x is the size parameter) starts to be larger than 1.
\begin{figure}
    \centering
    \includegraphics[width = \linewidth]{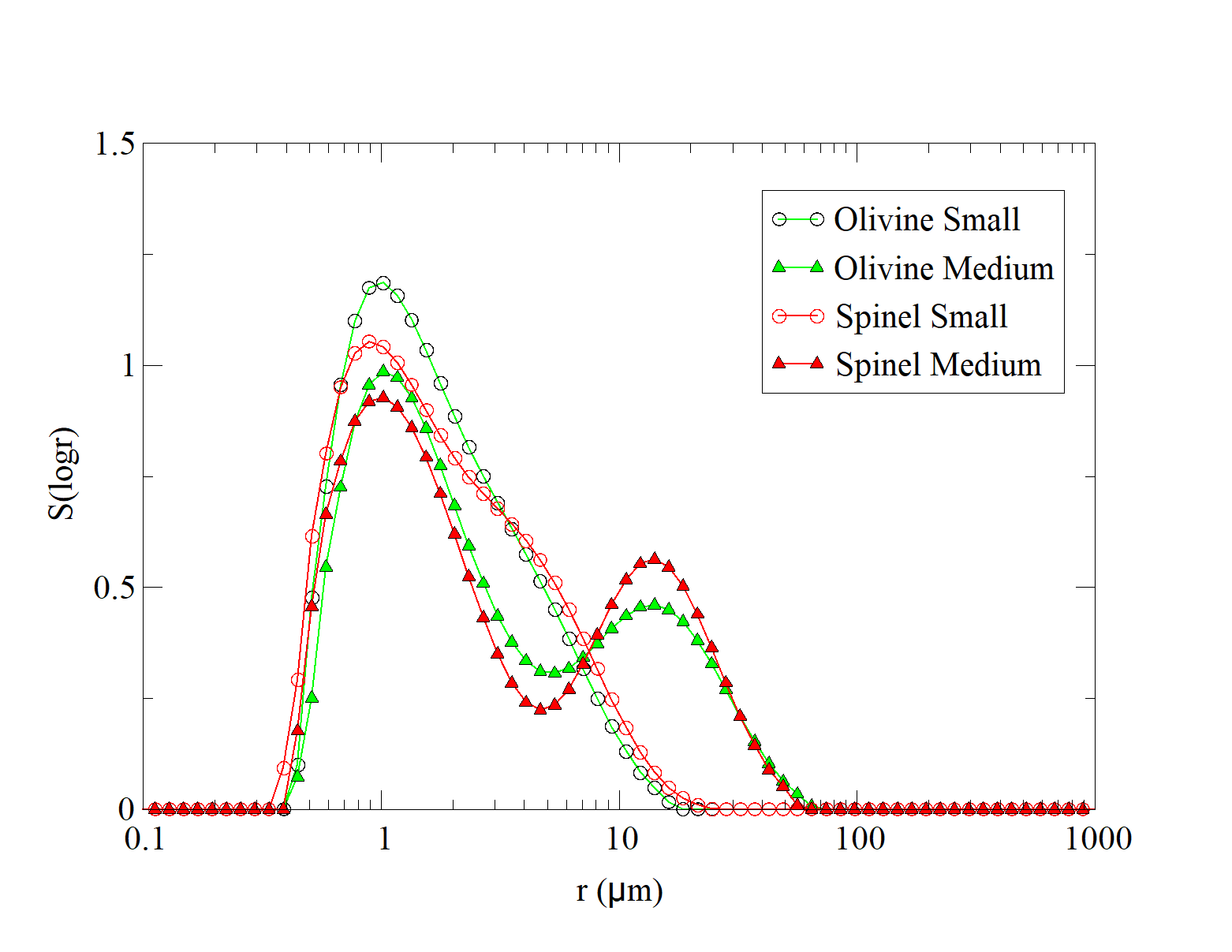}
    \caption{Size distribution of the four powdered samples.}
    \label{fig:sd}
\end{figure}

\subsection{Size distribution}
\label{sizes}
The particle size distributions (PSDs) of the powdered samples are obtained with a Laser Light Scattering (LLS) particle sizer (Malvern Mastersizer (2000); \cite{iso}). 
The LLS method is based on the measurement of the phase function of samples dispersed in a carrier fluid at $\lambda = 633$ nm within a range of low scattering angles (0.02$^{\circ}$–30$^{\circ}$) and a few larger scattering angles (45$^{\circ}$, 60$^{\circ}$, 120$^{\circ}$, 135$^{\circ}$).
The volume distribution of equivalent spherical particles that best reproduces the observed phase function is obtained by inverting a light scattering model based on    Mie theory, which requires knowing the complex refractive index of the samples (equation \ref{eq:index}).
From the retrieved volume size distributions we obtain the corresponding projected-surface-area distribution, $S(log r)$, of an equivalent projected surface sphere with radius $r$ (Figure \ref{fig:sd}).\\
All  PSDs show a well-defined primary peak around 1$\mu$m, while the \textit{Olivine Medium} and the \textit{Spinel Medium} show also a secondary peak around 20 $\mu$m, that extends up to 100 $\mu$m.  
The PSDs of the powered samples can be characterized by the effective radius $r_{\rm eff}$ and the effective variance $\sigma_{\rm eff}$ as defined by \cite{hansen}. 
These parameters have a direct interpretation for
mono-modal distributions, while for multimodal distributions, they are only first-order indicators  of   particle  size.
Table \ref{tab:size_distribution_parameters} shows the effective radius $r_{\rm eff}$, the effective variance $\sigma_{\rm eff}$ and the effective size parameter, i.e., $x_{\rm eff} = 2 \pi r_{\rm eff}/\lambda$ of the samples.  In the case of the 
\textit{Olivine Pebble} and  \textit{Spinel  Pebble}, their sizes are defined by
the radii of volume-equivalent spheres.

 \begin{table}
     \centering
     \begin{tabular}{lccc}
Sample  & $\rm r_{\rm eff}$ ($\mu$m) & $\sigma_{\rm eff}$ ($\mu$m) & $x_{\rm eff}$ \\
\hline\hline
Olivine Small   &  2.4   &  1.0  & 29 \\
Spinel Small    &  2.6  &  1.0  & 32  \\
Olivine Medium    &  6.5  &  1.4  & 79 \\
Spinel Medium   &  7.2 &  1.3  & 88 \\
\hline
Olivine Pebble &  3.8 mm$^{\rm a}$ &-- &  46$\cdot$ 10$^4$ \\
Spinel Pebble  &  3.4 mm$^{\rm a}$ &-- & 41$\cdot$ 10$^4$    \\
\hline\hline
     \end{tabular}
     \caption{Characteristic parameters of the size distribution. $\rm r_{\rm eff}$ is the effective radius, $\sigma_{\rm eff}$ is the effective variance and $x_{\rm eff}$ is the size parameter computed from the effective radius.$^{\rm a}$Radius of the volume-equivalent sphere.}
     \label{tab:size_distribution_parameters}
 \end{table}


 \begin{figure*}
    \centering
    \includegraphics[width = \linewidth]{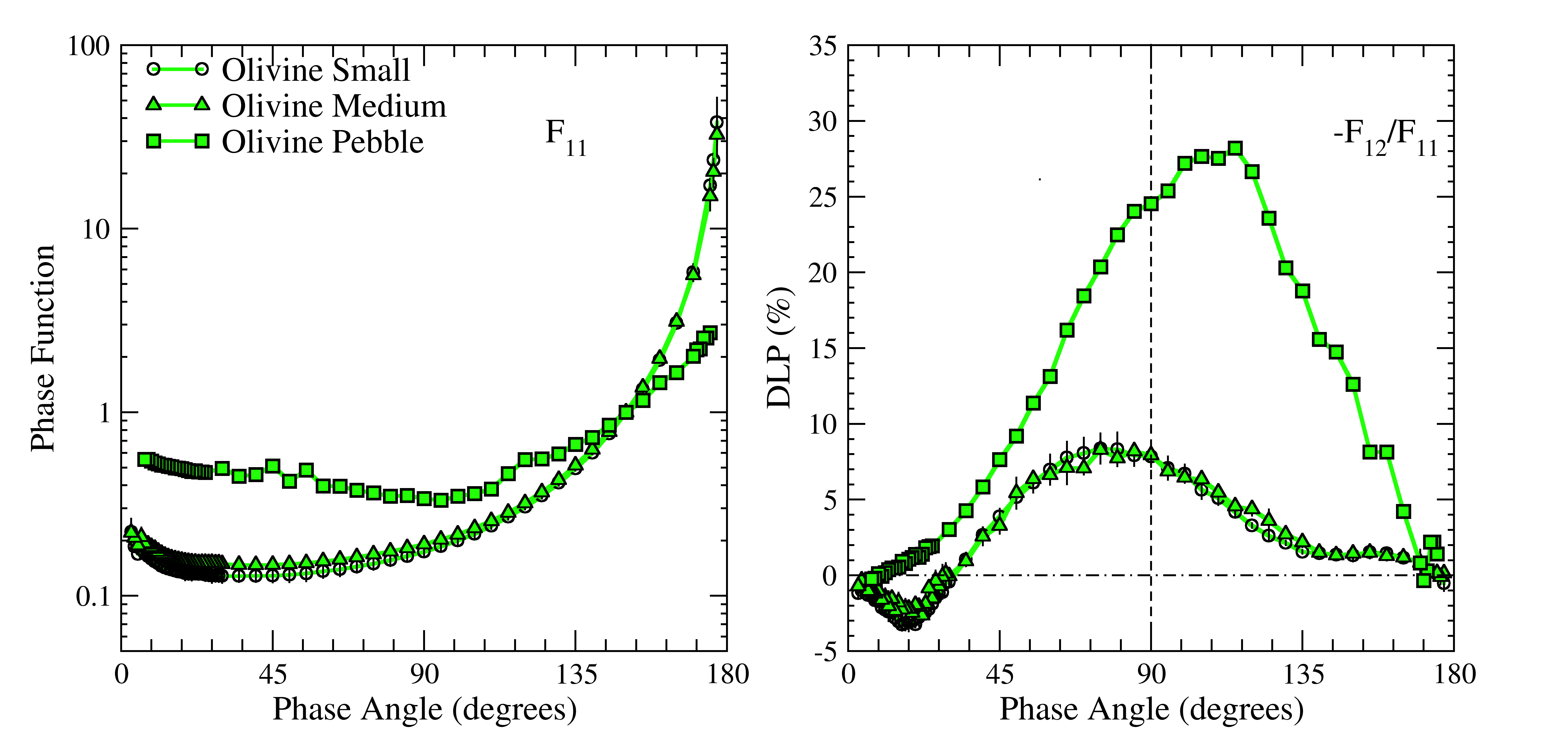}
    \caption{Phase function (left) and degree of linear polarization (right) of the olivine  samples. The phase functions are normalized to 1 at 150 degrees.}
    \label{fig:olivine_all}
\end{figure*}

 \begin{figure*}
    \centering
    \includegraphics[width = \linewidth]{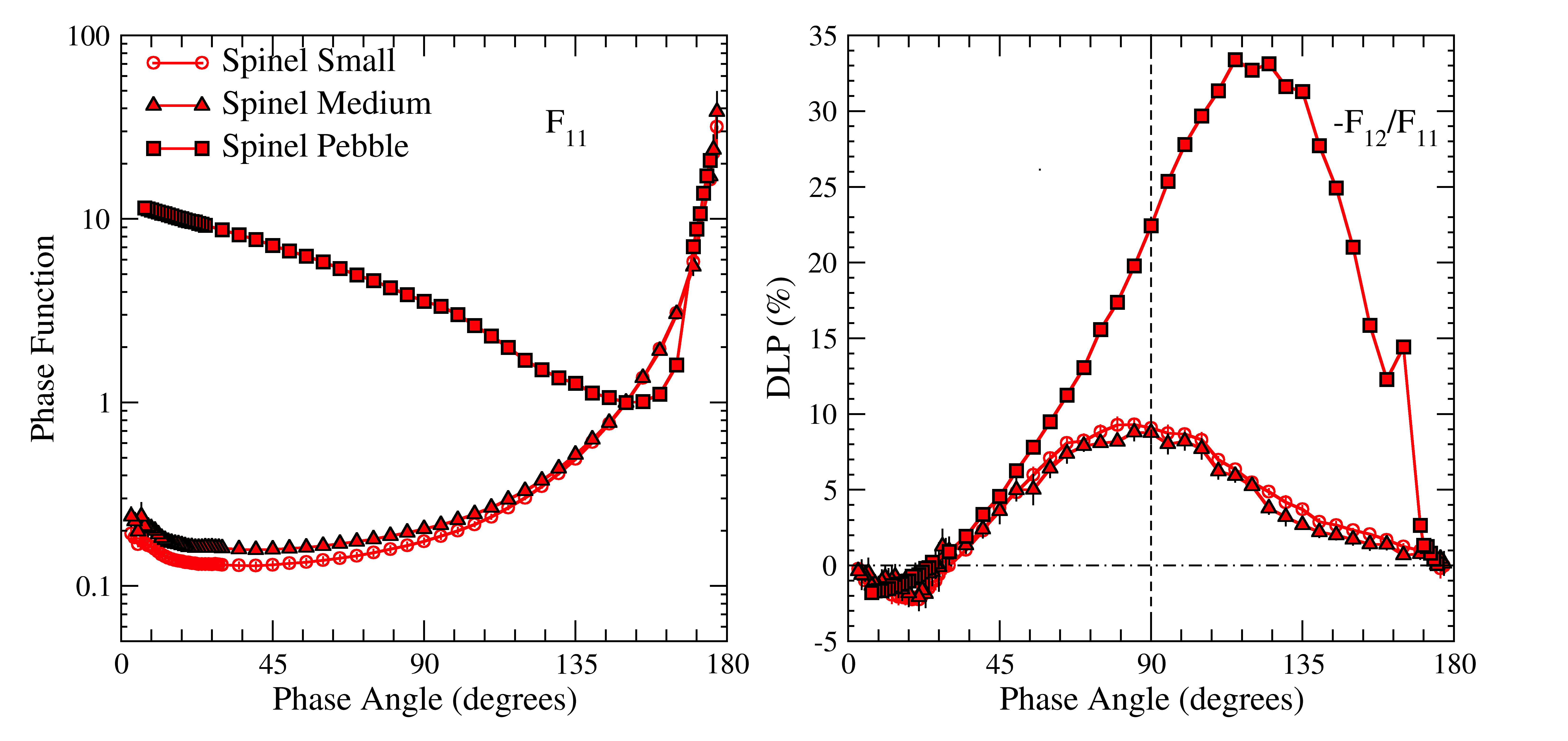}
    \caption{Phase function (left) and degree of linear polarization (right) of the spinel  samples. The phase functions are normalized to 1 at 150 degrees.}
    \label{fig:spinel_all}
\end{figure*}

\begin{figure*}
    \centering
    \includegraphics[width=\linewidth]{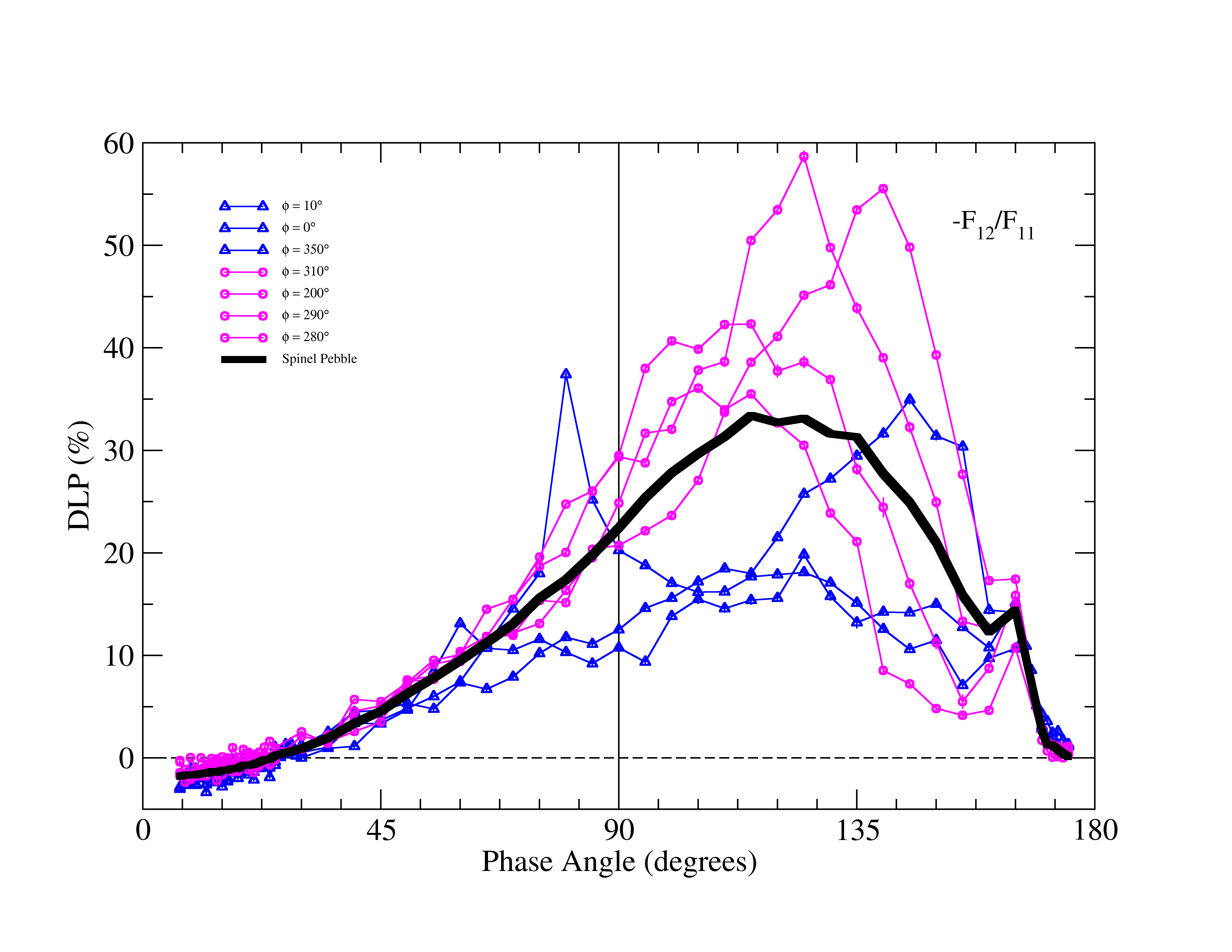}
    \caption{Phase-angle dependence of the degree of linear polarization of the \textit{Spinel Pebble}. Each curve corresponds to a different orientation of the grain. The blue triangles correspond to sharp-cornered sides, whereas the magenta circles to smooth faces. The black line is the average curve for the \textit{Spinel Pebble}. }
    \label{fig:spinel_orientation}
\end{figure*}

 \begin{figure*}
    \centering
    \includegraphics[width = \linewidth]{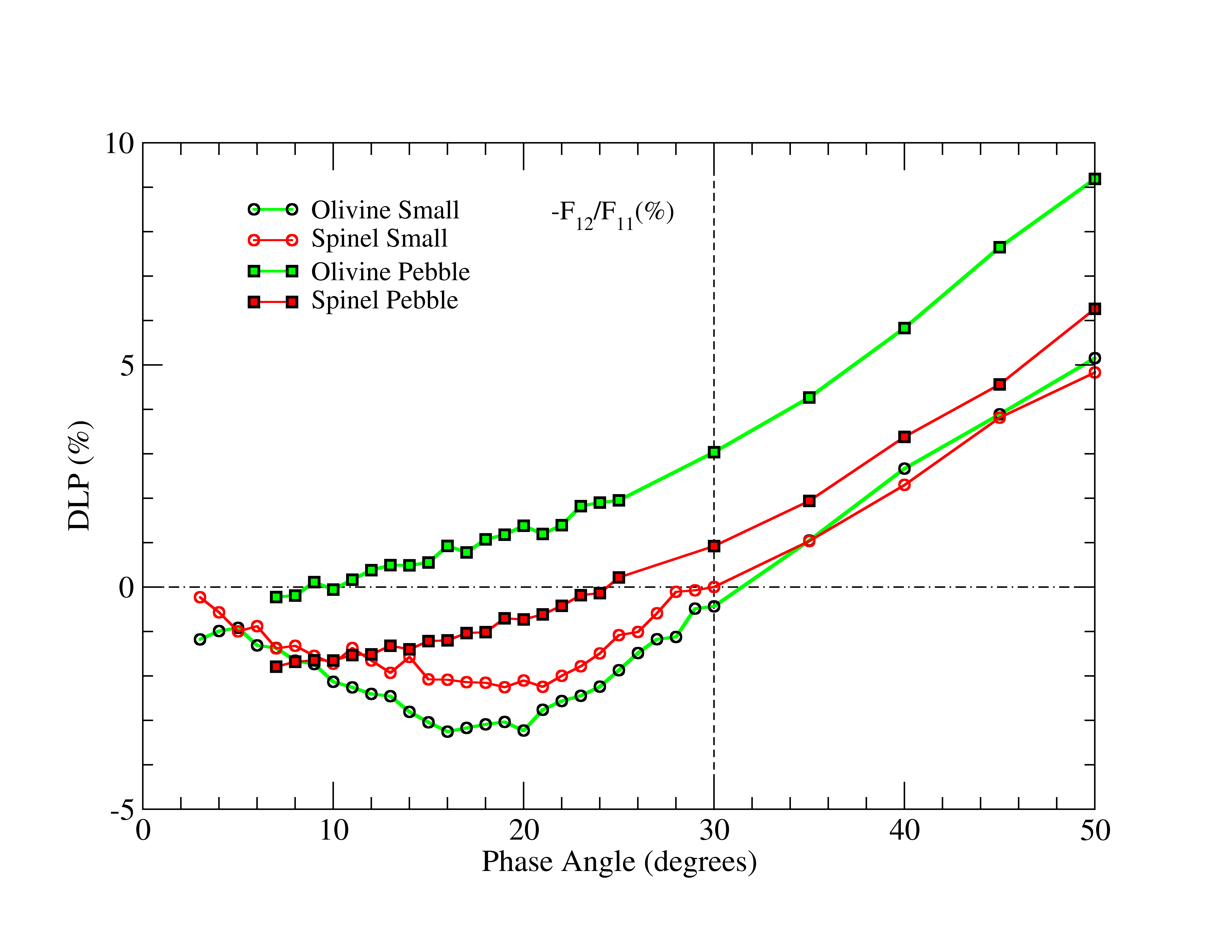}
    \caption{The figure shows the negative branch of linear polarization of the \textit{Olivine Small} and the \textit{Olivine Pebble}, in green, and of the \textit{Spinel Small} and the \textit{Spinel Pebble }, in red.}
    \label{fig:negative_branch}
\end{figure*}

\section{Measurements}
\label{measurements}
Figures \ref{fig:olivine_all} and \ref{fig:spinel_all} show the measured
normalized phase function $F_{11}(\alpha)/F_{11}(30^{\circ})$ and degree of linear polarization DLP$(\alpha) = -F_{12}(\alpha)/F_{11}(\alpha)$ for the olivine and spinel samples, respectively. 
The measured data of the powdered samples span from 3$^{\circ}$ to 177$^{\circ}$ in steps of 5$^{\circ}$ within the 30$^{\circ}$ -  175$^{\circ}$ phase angle range, and in steps of 1$^{\circ}$ within the 3$^{\circ}$ – 30$^{\circ}$ and 175$^{\circ}$ -  177$^{\circ}$ ranges.
For the  \textit{Pebble} samples, the measurements were carried out from 7$^{\circ}$ to 175$^{\circ}$ in steps of 5$^{\circ}$ within the 25$^{\circ}$ -  170$^{\circ}$ range, and in steps of 1$^{\circ}$ within the 7$^{\circ}$ – 25$^{\circ}$ and 170$^{\circ}$ -  175$^{\circ}$ ranges.
As explained in Section ~\ref{experimental_apparatus}, to simulate random orientation, the $F_{11}$ and $-F_ {12}/F_{11}$ curves, are obtained by  averaging $F_{11}^{\varphi}$ and $-(F_{12}/F_{11})^{\varphi}$  over 54 different orientations. Figure~\ref{fig:spinel_orientation} illustrates  the effect of rotation on the degree of linear polarization curves for the {\it Spinel Pebble}.  For simplicity, we only show a set of selected orientations for the {\it Spinel Pebble}. They are plotted together with the averaged values based on the 54 measured orientations. Figure~\ref{fig:spinel_orientation} shows the high dispersion of results for each of the individual orientations.

 \begin{table}
     \centering
     \begin{tabular}{lcccc}
Sample  & $\rm r_{\rm eff}$ ($\mu$m) & $\sigma_{\rm eff}$ ($\mu$m) & $f $ & BE\\
\hline\hline
Olivine Small   &  2.4   &  1.0  & 0.74     &   1.319 \\
Spinel Small    &  2.6  &   1.0  & 0.74    &  1.320 \\
Olivine Medium    & 6.5  &  1.4  & 0.77     &   1.313 \\
Spinel Medium   &  7.2 &  1.3  & 0.78     &   1.342 \\
\hline
Olivine Pebble    &  3.8 mm$^{\rm a}$ &  -  & 1.51     &   1.087 \\
Spinel Pebble  &   3.4 mm$^{\rm a}$  &   -      &   2.02  &   1.598\\
\hline\hline
     \end{tabular}
     \caption{Characteristic parameters of the phase function. $\rm r_{\rm eff}$ is the effective radius and $\sigma_{\rm eff}$ is the effective variance. $^{\rm a}$Radius of the volume-equivalent sphere. The parameter  $f = F_{11}(45^{\circ})/F_{11}(90^{\circ})/ $ is used to study the side phase-angles region, the parameter  $BE = F_{11}(7^{\circ})/F_{11}(45^{\circ}) $  evaluates the backscattering enhancement.}
     \label{tab:phase_function_parameters}
 \end{table} 
  
 \begin{figure*}
    \centering
    \includegraphics[width = \linewidth]{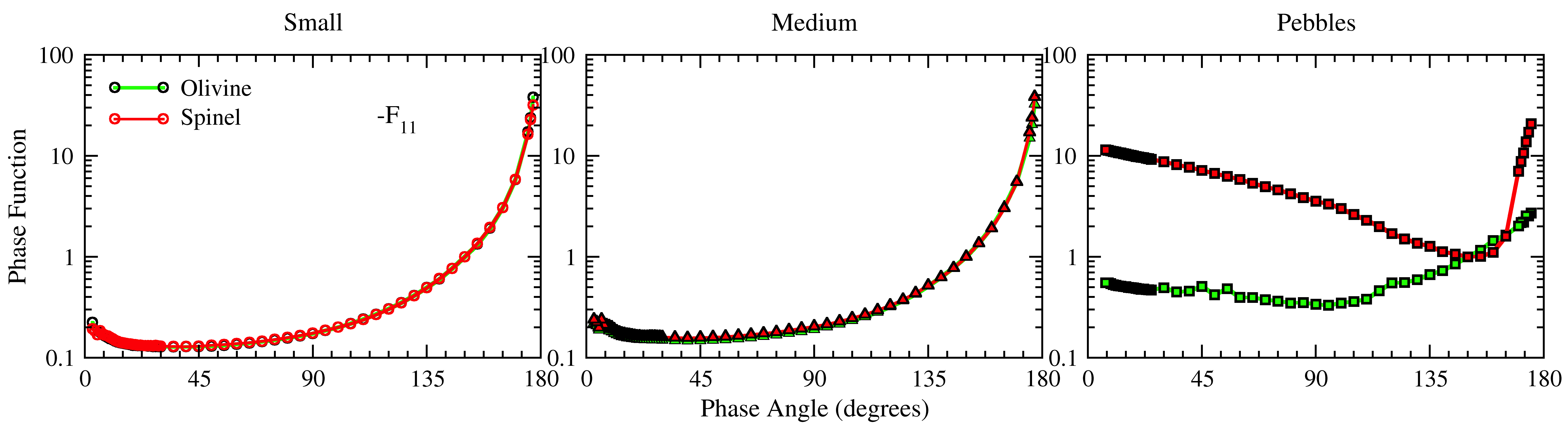}
    \caption{The three graphics represent respectively the phase function of the \textit{Small}, \textit{Medium}, and \textit{Pebble} samples  of olivine and spinel. All phase functions are normalized to 1 at 150 degrees.}
    \label{fig:phase_three_size}
\end{figure*}

 \begin{figure*}
    \centering
    \includegraphics[width = \linewidth]{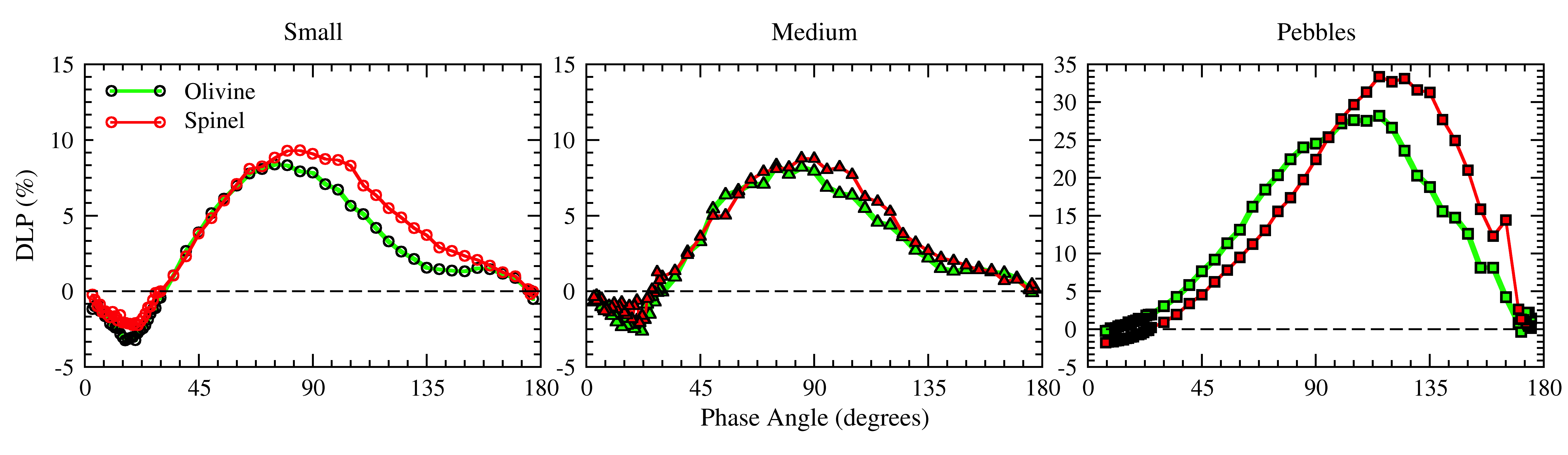}
    \caption{The three graphics represent respectively the DLP curves of the  \textit{Small}, \textit{Medium}, and \textit{Pebble} samples of olivine and spinel.}
    \label{fig:pol_three_size}
\end{figure*}

 \subsection{Size Effects}
The powdered and the pebble samples show strongly different scattering properties. 
The phase function curves (Figures \ref{fig:olivine_all} and \ref{fig:spinel_all}, left panel) of the \textit{Small} and \textit{Medium} powdered samples have a rather flat trend  at back- and side-phase angles and a strong increase in the forward direction.  
They are qualitatively similar to other samples consisting of micron-sized mineral particles investigated at CODULAB and available at the Granada-Amsterdam Light Scattering Database (see e.g. \cite{olga2000, volten2006, jesus2017, frattin2019, juan-carlos2021}).
In contrast, the \textit{Pebble} samples  show U-shaped phase functions with the minima located at phase angles $\sim$100$^{\circ}$  and  $\sim$160$^{\circ}$, respectively and monotonically increasing from the minimum towards backscattering.
We use the $f$ parameter, defined as $f = F_{11}(45^{\circ})/F_{11}(90^{\circ})$ to evaluate the flatness at  intermediate phase angles.
The closer to one the value of $f$ is, the flatter the curve.  
When $f$ is lower than 1, the phase function curve increases with the phase angle, as in the case of the powdered samples. The \textit{Small} and \textit{Medium} samples have a similar trend, with \textit{Medium} samples slightly flatter than \textit{Small} samples.
When $f$ is higher than 1, the curve decreases with the phase angle, as in the case of the mm-sized particles. The \textit{Spinel Pebble}  has the higher value of the  $f$ parameter and shows the steepest negative slope in this region.
Then, we use the BE parameter, defined as   $BE = F_{11}(7^{\circ})/F_{11}(45^{\circ})$ to evaluate the backscattering enhancement of the phase function curves. 
The \textit{Small} and \textit{Medium} samples have very similar values   indicating a moderate backscattering enhancement.
Table \ref{tab:phase_function_parameters} lists the $f$ and $BE$ values for all   samples.\\
 \begin{table*} 
    \centering
    \begin{tabular}{lccccccccc}
      Sample  & r$_{\rm eff}$ ($\mu$m) & $\sigma_{\rm eff}$ ($\mu$m)  & $x_{\rm eff}$ & DLP$_{min}$ (\%) & $\alpha_{min}$ (deg) &  $\alpha_{0}$ (deg) & DLP$_{max}$ (\%)& $\alpha_{max}$ (deg)\\
        \hline\hline 
Olivine Small  & 2.4   & 1.0  & 29 &-3.5$\pm0.4$ & 16$\pm1$  & 30$\pm1$ & 8.4$\pm$1  & 75$\pm5$   \\ 
Spinel Small     & 2.6  & 1.0   & 32  & -2.3$\pm$0.3  & 19$\pm1$   & 30$\pm1$ &  9.3$\pm$0.3 & 85$\pm5$  \\ 
Olivine Medium     & 6.5  & 1.4  & 79  & -2.6$\pm$0.4  & 22$\pm1$   & 30$\pm1$ &  8.4$\pm$1.0  & 75$\pm5$  \\ 
Spinel Medium   & 7.2 & 1.3  & 88 & -2.1$\pm$0.5  & 21$\pm1$   & 26$\pm1$ &  8.8$\pm$0.5 & 85$\pm5$  \\
\hline
Olivine Pebble     &   3.8 mm$^{\rm a}$  & -  &9.6$\cdot$10$^3$   & -  &   - & 10$\pm1$ &  28.2$\pm$39.7 & 115$\pm5$   \\ 
Spinel Pebble   & 3.4 mm$^{\rm a}$  &-  &12$\cdot$10$^4$  &  - & -  & 24$\pm1$ &  33.4$\pm$24 & 115$\pm5$  \\ 
        \hline\hline 
    \end{tabular}
    \caption{Polarimetric parameters of the samples.  r$_{\rm eff}$  and $\sigma_{\rm eff}$   are the effective radius and effective variance. $^{\rm a}$Radius of the volume-equivalent sphere. $P_{min}$ is the minimum of polarization and $\alpha_{min}$ the corresponding scattering angle. $P_{max}$ is the maximum of polarization at scattering angle $\alpha_{max}$ and  $\alpha_{0}$ is the inversion angle.   
}
    \label{tab:polarization_parameters}
\end{table*}

The DLP curves show the characteristic bell shape for irregular particles with a negative branch at small phase angles and a maximum at side-phase angles (Figures \ref{fig:olivine_all} and \ref{fig:spinel_all}, right panels).
The main effects of   size on DLP curves are the variation of the maximum of polarization and the change of the depth of the  negative branch at small phase angles \citep{olga2021}. 
Table \ref{tab:polarization_parameters} lists the characteristic parameters of the DLP curves, in the maximum (DLP$_{max}$,$\alpha_{max}$), minimum (DLP$_{min}$,$\alpha_{min}$), and inversion ($\alpha_0$) regions.\\
The \textit{Spinel Pebble} shows the highest polarization maximum (DLP$_{max}=33.4$\%), followed by the \textit{Olivine Pebble} (DLP$_{max}=28.2$\%).
We notice that the DLP$_{max}$ of the \textit{Pebble} samples is significantly shifted toward larger phase angles, $\alpha_{max}=115^{\circ}$, with respect to the powdered samples, $\alpha_{max}=75^{\circ}-85^{\circ}$. 
The deepest negative polarization branch is observed for the \textit{Olivine Small} (DLP$_{min}$  = 3.5\%), and tends to be shallower as  the size of the particles increases.\\
Figure \ref{fig:negative_branch} shows in detail the trend of the negative polarization branch for the \textit{Small}  and  \textit{Pebble} samples. 
The three spinel samples show a well defined negative polarization branch with a high inversion angle, $\alpha_0$, regardless of the particle size, which has values of 30$^{\circ}$ and 26$^{\circ}$ respectively for the \textit{Spinel Small} and the \textit{Spinel Medium} and value of 24$^{\circ}$, for \textit{Spinel Pebble}.
It is interesting to note that in the case of the olivine samples
the inversion angle is highly dependent on the particle size and it reaches the minimum value of $10^{\circ}$ for the \textit{Olivine Pebble}, when the negative branch almost disappears.\\

\subsection{Composition and Macroscopic Effects}
Figure \ref{fig:phase_three_size} shows the olivine and spinel phase functions, respectively for the \textit{Small}, \textit{Medium}, and \textit{Pebble} samples. Powdered samples of the same size show almost coincident curves, while pebbles behave differently, most likely due to  their shape and macroscopic structure, smooth and rounded for the olivine and sharp and multifaceted for the spinel.\\
Figure \ref{fig:pol_three_size} shows together the olivine and spinel DLP curves  for \textit{Small}, \textit{Medium}, and \textit{Pebble} samples.
The differences between curves for the two \textit{Small} samples may be attributted to the small differences in the refractive index of the olivine and spinel. Spinel is more absorbing than olivine because of its higher Fe and Cr content, which may explain its higher DLP$_{max}$ and shallower negative polarization branch (Table \ref{tab:polarization_parameters}).
The dip around 135$^{\circ}$ has been measured also for Mg-rich olivine by \cite{olga2000, olga2021}.\\
We notice that although the DLP curves of the \textit{Small} and \textit{Medium} samples display similar trends, the \textit{Medium} samples show   higher dispersion of values. 
A possible explanation of this effect lies in the fact that the \textit{Small} samples have mono-modal narrow size distribution, peaking at 1 $\mu$m, while the \textit{Medium} samples have a broader bi-modal size distribution, including larger particles. 
\cite{jesus2018} showed that, by removing particles smaller than 1 $\mu$m from a lunar dust analog sample, the negative polarization branch (NPB) nearly vanishes. This result has been recently confirmed by \cite{olga2021}.\\
 The two \textit{Pebble} samples produce quite different DLP curves, likely
related to their macroscopic structure.
Figure \ref{fig:spinel_orientation} shows the DLP curves of the \textit{Spinel Pebble} for different orientations $\varphi$ of the holder with respect to the incident laser beam, as described in Section \ref{experimental_apparatus}. 
The blue triangles indicate the curves generated when the sharp-cornered sides face the laser beam, whereas the magenta circles denote the smooth sides. Interestingly, the  NPB appears for all pebble orientations and it is deeper when the flat faces with sharp corners are facing the beam. This indicates that internal reflections from troughs or corners with right interfacial angles might be responsible for the measured NPB (see Section \ref{observations}).\\

 \begin{figure*}
    \centering
    \includegraphics[width = \linewidth]{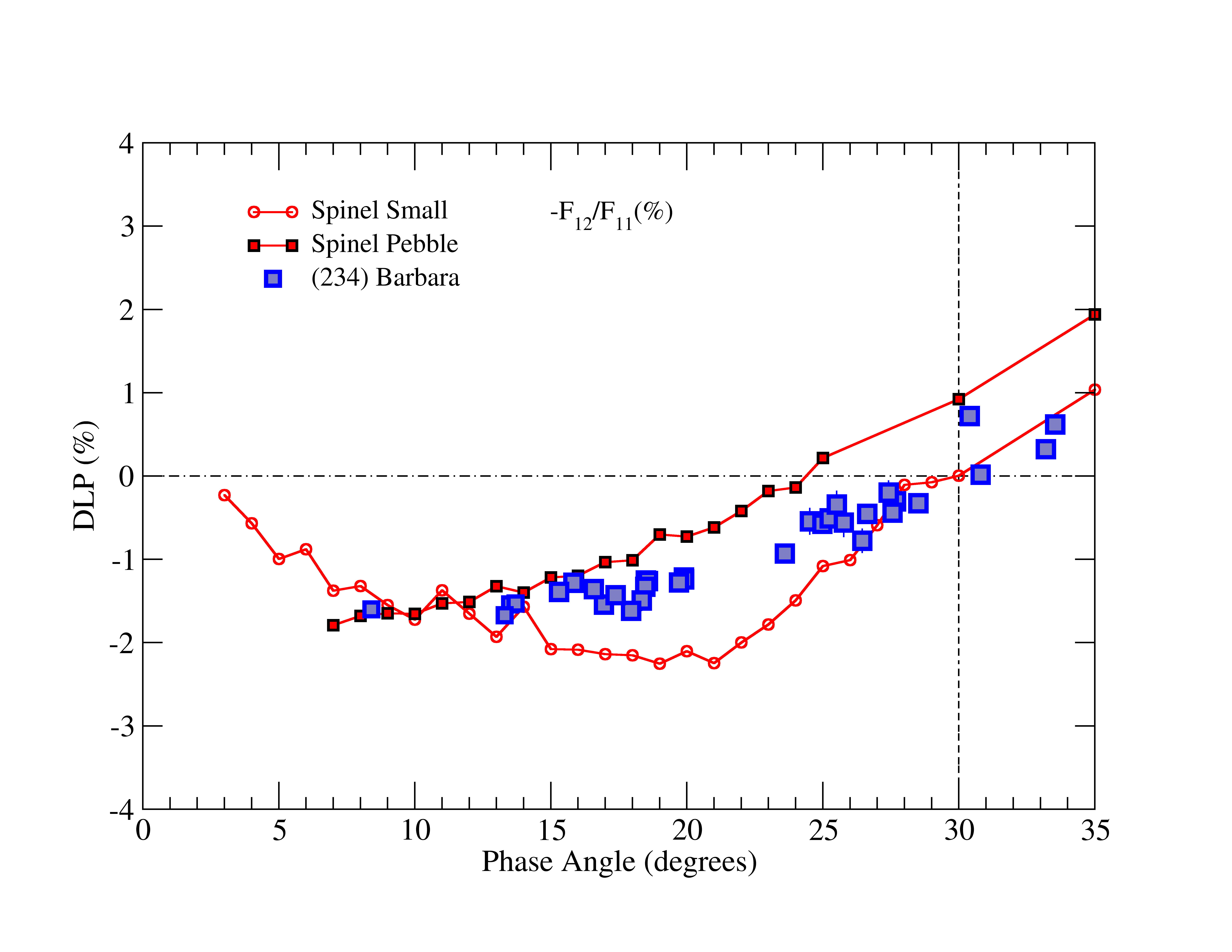}
    \caption{The figure shows the negative branch of linear polarization of the \textit{Spinel Small}, in empty red circles, and the \textit{Spinel Pebble}, in full red squares, together with the (234) Barbara asteroid values of polarization, in blue squares, measured by \citet{cellino2006,gil-hutton2008,masiero2009,devogele2018}.}
    \label{fig:negative_branch_Barbara}
\end{figure*}

\section{Comparison with Barbarians} 
\label{observations}

The NPB is a well-known feature in the polarization curves of atmosphereless bodies and planets of the solar system (i.e.,  asteroids \citep{fornasier2006}, Galilean satellites of Jupiter \citep{rosenbush1997}, Mars \citep{shkuratov2005}, etc.).
When the light of the Sun is scattered by the surface of a body (e.g., an asteroid), it becomes partially linearly polarized and can be characterized by its parallel, P$_{\parallel}$, and perpendicular, P$_{\perp}$, components.  The NPB arises when the polarization plane is parallel to the scattering plane. Conversely, positive polarization occurs when the polarization plane  is perpendicular to the scattering plane. The appearance of this feature has been extensively investigated through laboratory experiments  \citep{shkuratov2006, hadamcik2006, frattin2019, olga2021} and theoretical simulations \citep{muinonen2012, liu2015, jesus2017, Huang2020}. The NPB is explained by the spatial assymmetry of the internal fields of  a irregular wavelength-scale single-particle (single-particle mechanism) \citep{muinonen2011}, and by the so-called coherent backscattering mechanism (CBM), a phenomenon in which the  radiation reaches a maximum in the backward direction due to the interference of the scattered light beam produced by the single particles of a cloud \citep{muinonen1989,shkuratov1989,muinonen1990,muinonen2012}.

For mm-sized particles in the geometric optics regime, there is evidence that the   presence of flat surfaces at right angles (i.e. the particles are crystals) leads to retro-reflection that would cause a NPB. The NPB is deeper when the  interfacial angle approaches  90$^{\circ}$ \citep{muinonenetal1989}.
These mechanisms may overlap to different extents in complex media such as asteroidal regoliths.
The \textit{Small} and \textit{Medium} samples of olivine and spinel show similar light scattering properties, and the differences result most likely from the higher absorption coefficient of spinel. 
In contrast, the \textit{Pebble} samples   have   different macroscopic shapes.
Spinel shows a multifaceted surface with a certain degree of roughness and   olivine   is a rounded peridot with a smooth surface.
Therefore, it is likely that the macroscopic structure
is involved in the development of the NPB of the mm-sized pebbles in the geometric optics regime.
Previous work has shown that a deep negative branch is also expected for  orthorhombic  forsterite crystals \citep{muinonenetal1989}.
The new experimental evidence that mm-sized spinel particle shows a NPB could help in the interpretation of the NPB in the Barbarians.
This class of asteroids has a peculiar polarimetric behavior, characterized by an  extended NPB, with a large inversion angle.  
\cite{devogele2018} and \cite{sunshine2008} proved that Barbarians belong to class L  (DM taxonomy, \cite{demeo2009}), characterized   by an absorption band around 2 $\mu$m generated by the FeO-rich spinel in the CAIs. 
Figure \ref{fig:negative_branch_Barbara} shows the NPB of the \textit{Spinel Pebble} and \textit{Spinel Small} together with the value of polarization retrieved by \cite{cellino2006,gil-hutton2008,masiero2009,devogele2018} for the asteroid (234) Barbara. 
It can be seen that the \textit{Spinel Pebble} NPB shows a   similar trend to that of the Barbara asteroid, with a very high inversion angle and a low $\alpha_{min}$. 
The DLP curve of  \textit{Spinel Small} (and of  \textit{Olivine Small}) also shows a high inversion angle, but forms a well defined minimum around a high $\alpha_{min}$ ($\sim$20$^{\circ}$). An extended NPB with a minimum very close to the forward direction is a characteristic feature of the DLP curve of crystals larger than the wavelength of the incident radiation \citep{muinonenetal1989}.
Since the discovery of the peculiar class of Barbarians asteroids, various attempts have been done in order to explain their large inversion angle. \cite{cellino2006} and \cite{delbo2009} hypothesized the large concavities  due to impacts and large scale craters on the surfaces could be   responsible for the large inversion angle. \cite{gil-hutton2008} proposed that the feature was due to a mix of high and low albedo particles, and \cite{devogele2018} suggested that this behavior could be related to the unusually small size  of the particles forming the Barbarians  surface regolith.\\
Asteroidal regolith is composed of a variety of materials with different compositions, optical properties, and size distributions, as shown in situ by the Hayabusa and OSIRIS-REX missions \citep{hamilton2021, cambioni2021}. Previous work has favoured scattering by micron-sized particles as an explanation of the opposition effect in the phase function and the negative branch in the polarization curves of asteroids. Single micron-sized particles give rise to gentle negative polarization and subtle increase in brightness towards backscattering, whereas large systems of such particles, due to the coherent backscattering mechanism, give rise to sharp opposition effects and negative polarization features closer to the backward scattering direction (e.g., \cite{muinonen1990, shkuratov1994, grynko2022}). Our results  suggest that the size distribution of the surface regolith particles of the Barbarians may be shifted toward larger sizes compared to other asteroidal families, resulting in a significant contribution of geometric optics retro-reflection to the polarization curve.
The specific composition of spinel is unlikely to be decisive in determining the shape of the phase function and the DLP curves of Barbarian asteroids. Note that the amount of spinel in these objects could be relatively small. However, further analysis is required to understand which physical properties of the regolith materials are responsible for the polarization features and phase function curves of the Barbarian asteroids.

\section{Summary and Conclusions}
\label{conclusions}

In this work, we have measured the phase function and degree of linear polarization of six samples of olivine and spinel with different sizes.
The \textit{Small} and \textit{Medium} samples are composed of micron-sized irregular particles and they show a light scattering behavior qualitatively similar   to that of    other powdered samples with similar size distributions. 
The \textit{Pebble} samples consist of mm-sized particles and they have more variable and complex behavior.
The phase function of  the \textit{Small} and \textit{Medium} samples is flat with a strong increase in the forward direction, both for olivine and spinel, whereas the larger  \textit{Pebble} samples show a clear U-shape curve. 
The degree of linear polarization curves of the  \textit{Small} and  \textit{Medium} samples show the characteristic bell shape with a negative branch at small phase angles and a maximum at side-phase angles. 
The  \textit{Pebble} samples show the DLP$_{max}$ shifted toward larger phase angles. 
The \textit{Spinel Pebble} shows a clear negative polarization branch
whereas the \textit{Olivine Pebble} does not.
Currently, the single-particle and  coherent back-scattering mechanisms arethe forefront explanation of the NPB of clouds and regoliths composed of wavelength-scale particles.  The case of the {\it Spinel Pebble}, the lack of surface roughness suggests that retro-reflection generated by the sharp corners may be responsible for the measured DLP. Therefore both size and macroscopic structure of the samples may play a role in the production of the NPB. \\
In the astrophysical domain, the peculiar class of Barbarian asteroids shows a NPB with a very large inversion angle.
In this work, we show that the macrocospic shape of the \textit{Spinel Pebble} can generate a NPB with a large inversion angle. The experimental data presented in this work can also be of interest for interpretating the polarimetric behaviour of the olivine-rich A-class asteroids. In the future, we plan to experimentally investigate the scattering properties of other components of CV3-like meteorites, representative of the Barbarians composition, to check their contribution to the global polarization of the targets.
  
 
Data Availability: The experimental data will be available at the Granada-Amsterdam Light Scattering Database (www.iaa.es/scattering). 

\section*{Acknowledgements}
 We are grateful to Ricardo Gil-Hutton for his fruitful review on a previous version of the manuscript.  This work has been funded by the Spanish State Research Agency and 
Junta de Andaluc\'{i}a through grants LEONIDAS (RTI2018-095330-B-100) and 
P18-RT-1854, and the {\it Center of Excellence Severo Ochoa} award to the 
Instituto de Astrof\'{i}sica de Andaluc\'{i}a (SEV-2017-0709). JC 
G\'{o}mez Mart\'{i}n and T Jardiel acknowledge financial support from the European Science Foundation (ESF) and the Ram\'{o}n y Cajal Program of MICINN. Research by KM supported by the Academy of Finland grants No. 345115 and 336546.


\bibliographystyle{mnras}
\bibliography{example} 







\bsp	
\label{lastpage}
\end{document}